\documentstyle[sprocl,picins,psfig]{article}

\def\be{\begin{equation}}
\def\ee{\end{equation}}
\def\bea{\begin{eqnarray}}
\def\eea{\end{eqnarray}}
\def\etal{{\it et al.\/}}
\begin{document}

\title{REAL COMPTON SCATTERING FROM THE PROTON}

\author{Alan M. NATHAN}

\address{University of Illinois at Urbana-Champaign, 1110 W. Green Street,\\
Urbana, IL 61801, USA\\E-mail: a-nathan@uiuc.edu}


\maketitle\abstracts{
Real Compton Scattering on the proton in the hard scattering regime
is investigated.  Recent theoretical developments are reviewed.  Plans
for new experimental studies at Jefferson Lab are presented.}

\section{Introduction} \label{sec:intro}
Real Compton Scattering (RCS) in the hard scattering limit is a powerful
probe of the short-distance structure of the nucleon.  It is a natural
complement to other hard exclusive reactions that are currently being
pursued at Jefferson Laboratory (JLab) and elsewhere, such as high Q$^2$ elastic form factors,
hard pion electroproduction and photoproduction, and Virtual Compton
Scattering (VCS), and Deeply Virtual Compton Scattering (DVCS).  The common
feature of these reactions is a hard energy scale, leading to the
factorization of the scattering amplitude into a part involving a hard
perturbative scattering amplitude, which describes the coupling of the
external particles to the active quarks, and the overlap of soft nonperturbative
wave functions, which describes the coupling of the active quarks to the
proton.  For RCS, the hard scale is achieved when the Mandelstam variables
$s$, $-t$, and $-u$ are all large, or equivalently when p$_\perp$ is large,
on the hadronic scale.

There has been considerable theoretical effort in recent years 
in calculating RCS cross sections and polarization observables in the
hard scattering limit, and one goal of this contribution is to reveiw
this activity (Section~\ref{sec:theory}).
Despite this renewed theoretical interest in RCS, the only
Compton scattering data available in this kinematic regime are the 20-year
old Cornell data,\cite{CornellCS} which are spase and of limited
statistical precision 
in the theoretically interesting range of high p$_\perp$.
In order to provide the high quality data necessary to discriminate among
reaction mechanisms and gain new insight into the structure of the
proton, a new experiment is planned at
JLab (E97-108/E99-114) to measure RCS cross sections over a broad range 
of $s$ and $t$
and to have an initial look at polarization observables.  A second goal
of this contribution is to present an overview of the new experiment
in the
context of the theoretical work (Section~\ref{sec:expt}).  A summary is
presented in Section~\ref{sec:summary}.

\section{Theoretical Overview} \label{sec:theory}

\begin{figure}[h]
\psfig{figure=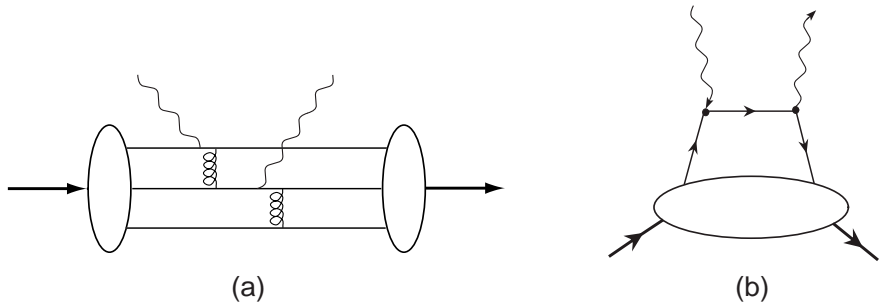,width=\textwidth}
\caption{Different hard scattering mechanisms for RCS.
In the pQCD mechanism (a), the
momentum is shared among the quarks by hard gluon exchange.  In
the handbag mechanism (b), the scattering is from a single quark
and the momentum is shared by the overlap of the high momentum
components of the soft wave function.}
\label{fig:hsa}
\end{figure}

Various theoretical approaches have been applied to RCS in the hard
scattering regime,
and these can be distinguished by the number of active
quarks participating in the hard scattering subprocess, or equivalently, by
the mechanism for sharing the
transferred momentum among the constituents.  Two extreme pictures
have emerged.
In the perturbative QCD (pQCD) approach,\cite{Kronfeld,Farrar,Vanderhaeghen}
there are three active quarks which share $t$
by the exchange
of two hard gluons (see Fig.~\ref{fig:hsa}a).  
In the soft overlap approach,\cite{Radyushkin1,Diehl} the handbag diagram
dominates (see Fig.~\ref{fig:hsa}b) in which
there is one active quark and $t$ is shared by the overlap of the
high momentum components of the soft wave function.  In any given kinematic
regime, both mechanisms will contribute, in principle, to the cross section.
It is generally believed that at sufficiently high energies, the pQCD 
mechanism dominates.  However,
the question of how high is ``sufficiently high'' 
is still an open question to be answered by more precise experiments, and
it is not known with any certainty what is the dominant mechanism in the
kinematic regime appropriate to JLab (p$_\perp \sim 1-2$ GeV).

\piccaptioninside
\piccaption{\label{fig:scaling}
Scaling of RCS
cross section at fixed $\theta_{cm}$.  The closed points are the
Cornell data. The open points represent the projected precision
from the JLab experiment. The curve is the prediction of
Radyushkin, assuming dominance of the handbag diagram.  For the
pQCD mechanism, n=6 independent of $\theta_{cm}$.}
\picskip{5}
\parpic[l]{\fbox{\psfig{figure=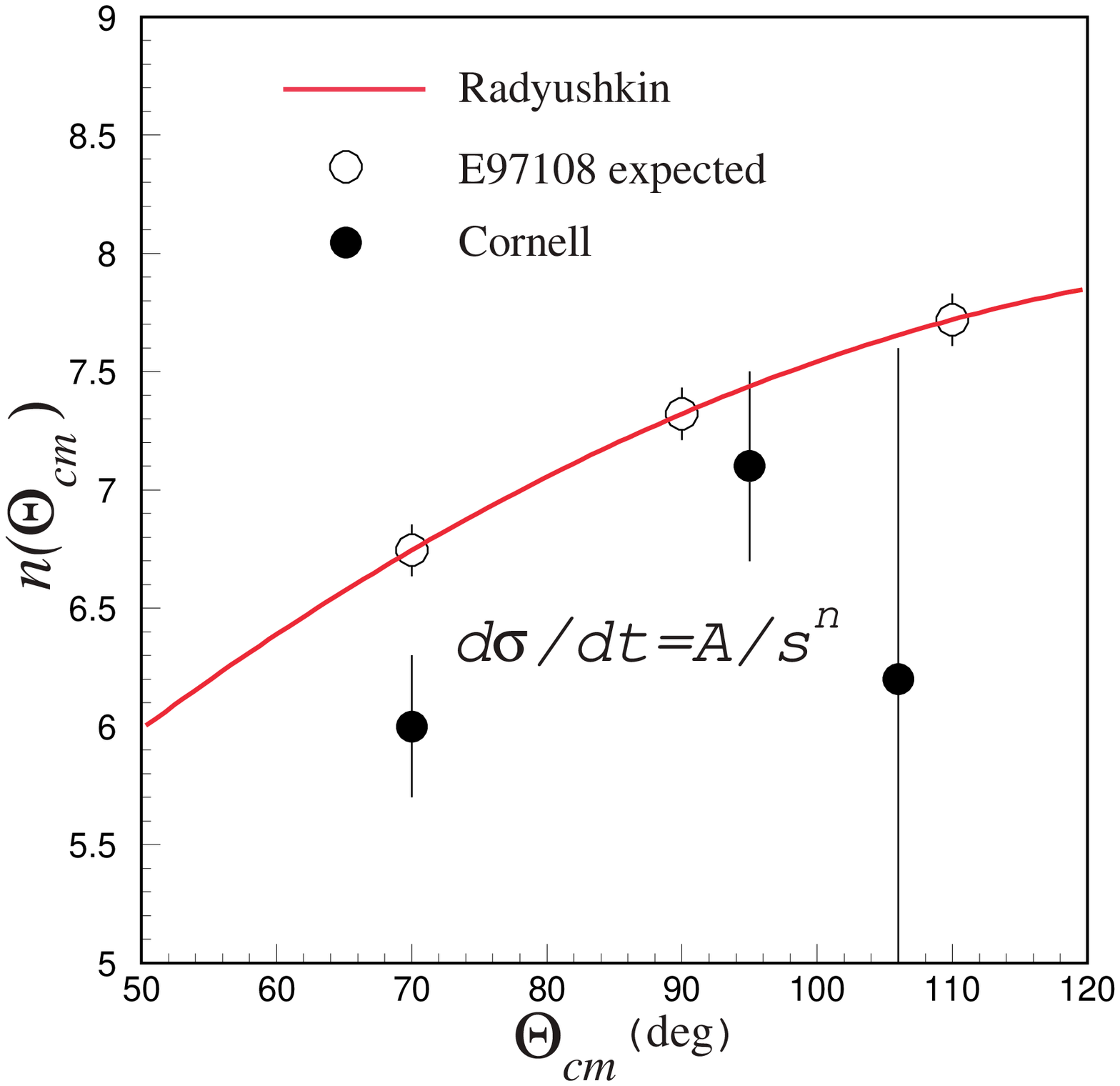,width=5cm}}}
I next examine the two reaction mechanisms in somewhat more detail.  In
the pQCD approach, the exchange of two hard gluons leads naturally to
the constituent counting rule and scaling \cite{BrodskyFarrar}
\begin{eqnarray}
{d\sigma\over dt} &=& \frac{f(\theta_{cm})}{s^n},
\label{eq:QuarkCounting}
\end{eqnarray}
where n=6 for RCS.  The valence Fock state dominates, since higher Fock states
require additional gluon exchanges and are therefore suppressed by additional
factors of $1/s$.  
The soft physics enters through
the valence quark distribution amplitude (DA) $\phi(x_1,x_2,x_3)$.
Experimentally, the Cornell data support scaling with $n\approx 6$
(Fig.~\ref{fig:scaling}), albeit with modest statistical precision.  
Nevertheless, it has been shown that the
cross sections calculated using the asympotic DA badly underpredict the
Cornell data.\cite{Kronfeld,Vanderhaeghen}   This has led to the suggestion \cite{Radyushkin1,Diehl} that the
dominant mechanism at experimentally accessible energies is the soft
overlap mechanism, which I describe next.

In the soft overlap approach to hard exclusive reactions in general and to RCS in
particular, the handbag diagram dominates.  The hard physics is contained in
the scattering from a single active quark, whereas the soft physics is contained
in the wave function describing how the active quark couples to the proton.
This coupling is described in terms of nonforward parton densities
(ND) \cite{Ji,Radyushkin2} which are superstructure of the nucleon
from which are derived the normal parton densities, elastic
form factors, and other quantities that have yet to be measured,
including new form factors accessible through Compton scattering.  Therefore
the ND provide links among diverse physical processes, including both
inclusive and exclusive reactions.
For example, the dominant ND in RCS is ${\cal F}^a(x,t)$, which is related
to both the Dirac form factor F$_1$ and the RCS vector
form factor R$_V$ through the expressions
\begin{equation}
F_1(t) \,=\, \Sigma_a e_a\int_0^1{\cal F}^a(x;t)dx
\phantom{space}
R_V(t)\, =\, \Sigma_a e^2_a\int_0^1{\cal F}^a(x;t)\frac{dx}{x} \, ,
\label{eq:ND}
\end{equation}
whereas the $t=0$ limit of ${\cal F}^a(x,t)$ is the parton distribution 
function $q^a(x)$.  Additional links are shown in
Table~1.
Despite the similarity between the $(e,e)$ and RCS form factors, an important distinction is the weighting
by the quark charge, which is linear for $(e,e)$ and quadratic for RCS, reflecting
the one- and two-photon nature of the interaction, respectively.  
Thus RCS is sensitive to the flavor structure of the proton in a different
way than electro-weak scattering, thereby potentially providing
another tool, along with parity-violating electron scattering, for
decomposing the flavor structure.  In particular, RCS has a greater sensitivity
to sea quarks than does $(e,e)$. 
\begin{table}[hpt]\label{tab:ff}
\caption{Nonforward parton densities (ND) and their associated
electroweak scattering (EWS) and RCS form factors, where $a$ labels
the quark flavor.  The last column shows
the relationship with the parton distribution functions.}
\vspace{0.2cm}
\begin{center}
%\footnotesize
\begin{tabular}{|c|c|c|c|}
\hline
ND & EWS Form Factor & RCS Form Factor	&	$t=0$ limit \\ \hline
${\cal F}^a(x;t)$ & F$_1(t)$	&	R$_V(t)$	&	$q^a(x)$	 \\
${\cal K}^a(x;t)$ & F$_2(t)$	&	R$_T(t)$	&	$2J^a(x)/x-q^a(x)$ 	 \\
${\cal G}^a(x;t)$ & G$_A(t)$	&	R$_A(t)$ 	&	$\Delta q^a(x)$	 \\
\hline
\end{tabular}
\end{center}
\end{table}

\begin{figure}[hp]
\psfig{figure=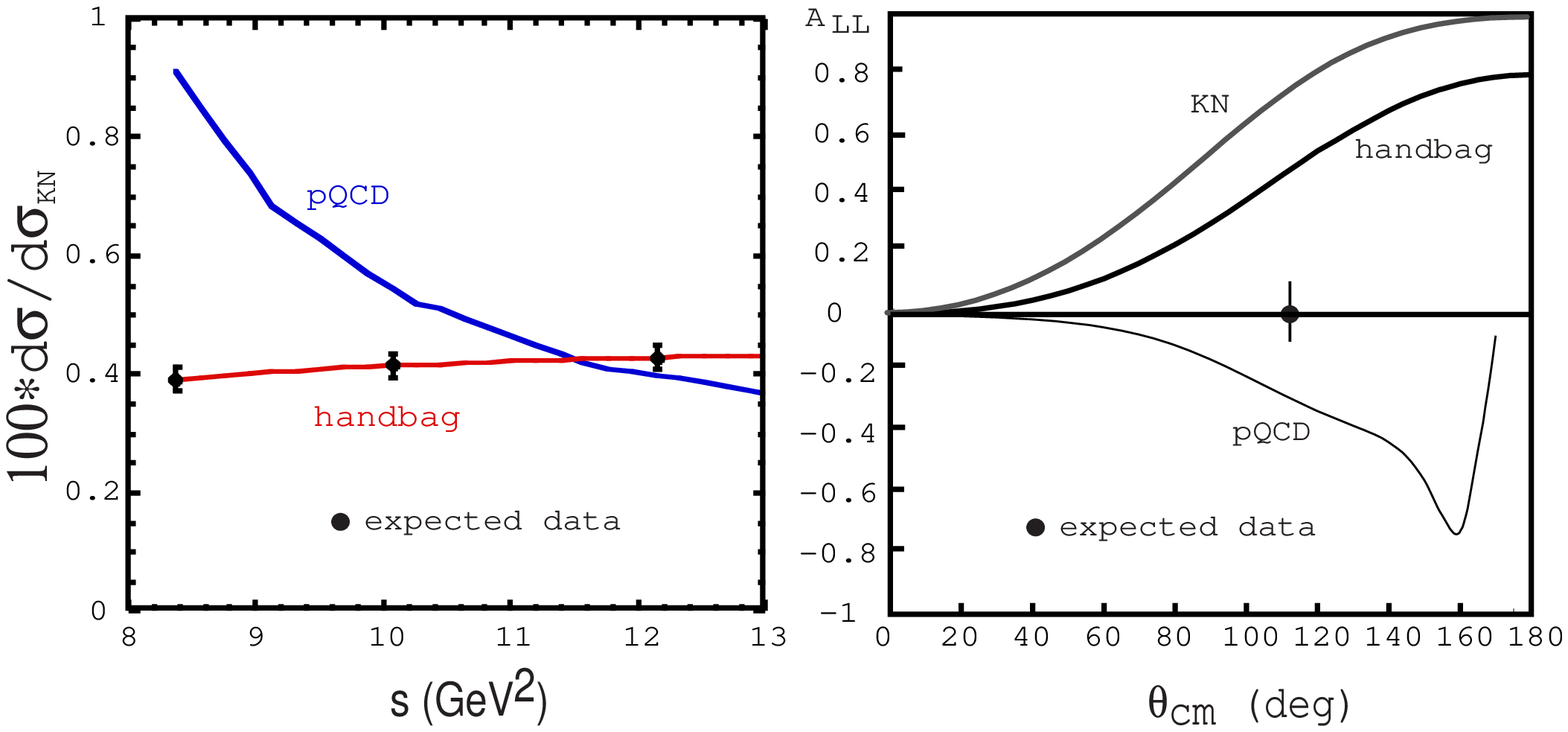,width=\textwidth}
\caption{(Left) The ratio $d \sigma$/$d \sigma_{\rm KN}$, 
scaled by a factor of 100,
as a function of $s$ at $-t=5$~GeV$^2$.  This
ratio is nearly independent of $s$ in the handbag but not in the
pQCD model. (Right) The longitudinal spin transfer parameter at 4 GeV.
The handbag model essentially follows the Klein-Nishina (KN) result for
a structureless Dirac particle.  For both plots the points show the precision
of the data expected from JLab E99-114.}
\label{fig:pseudo}
\end{figure}

The relationship between the RCS cross section and the form factors
has been worked out with several simplfying approximations, including
the neglect of terms such as R$_T$ which involve
hadron helicity flip.  This leads to the factorization of the RCS cross
sections into a simple product of the Klein-Nishina (KN) cross section
describing the hard scattering from a single quark and a sum of
form factors depending only on $t$ \cite{Radyushkin1,Diehl}:
\begin{equation}
\frac{d\sigma}{d\sigma_{\rm KN}}\,=\,f_V R_V^2(t) + (1-f_V) R_A^2(t)
\phantom{space}
f_V\,=\,\frac{(\tilde s-\tilde u)^2}{2(\tilde s^2+\tilde u^2)} \, ,
\label{eq:kn} 
\end{equation} 
where $\tilde s$ = $s-m^2$ and $\tilde u$ = $u-m^2$.  
The vector and axial vector form factors $R_V$ and $R_A$,
respectively,  have a simple physical
interpretation. The combination $|R_V(t)+R_A(t)|^2$ is the
probability that a photon can scatter elastically from the proton
by transferring $t$ to a single active quark whose helicity is
oriented in the direction of the proton helicity.  Similarly
$|R_V(t)-R_A(t)|^2$ is the probability that the active quark has
helicity opposite to that of the proton.
In order to measure $R_V$ and $R_A$, it is necessary to measure
the RCS cross section at fixed $t$ with a variable $f_V$ in order
to achieve a ``Rosenbluth-like'' separation. 
However, for the kinematics of interest,
where $s$, $-t$, and $-u$ are all large,  $f_V$ is always close to
1.  Consequently the unpolarized cross sections are largely
insensitive to $R_A$.  This leads to the very nice feature that
the left-hand-side of Eq.~\ref{eq:kn} is nearly $s$-independent at
fixed $t$.  This is shown in Fig.~\ref{fig:pseudo}, where one sees that
the pQCD mechanism predicts a very different behavior, thereby allowing a
very powerful test of the reaction mechanism as well as a precise measurement
of $R_V$.

Simple models for the ND's have been proposed, leading to predictions for
the RCS form factors and cross sections.\cite{Radyushkin1,Diehl}  An
interesting feature of these form factors is that they decrease approximately
as $1/t^2$ in the few-10 GeV$^2$ range, leading to $n\approx 6$ scaling
factor, in agreement with the asymptotic law (Eq.~\ref{eq:QuarkCounting})
but for very different reasons.  However, nontrivial violations of $n=6$
scaling are predicted in the form of an angle-dependent scaling
factor $n(\theta_{CM})$, which seems to agree with the limited Cornell
data (Fig.~\ref{fig:scaling}).  Another interesting feature is that at
sufficiently high $-t$, the form factors fall as $1/t^4$, leading to
$n\approx 10$.  Thus the handbag contribution to
RCS will be asymptotically subdominant to the pQCD contribution, even though
the former may dominate at experimentally accessible energies.

A measurement of polarization
observables provides further tests of the reaction mechanism as
well as access to additional form factors.  
The longitudinal polarization transfer observable $A_{LL}$ is
defined by 
\begin{equation} 
A_{LL}\frac{d\sigma}{dt}\,\equiv
\,\frac{d\sigma(\uparrow\uparrow)}{dt}-
\frac{d\sigma(\uparrow\downarrow)}{dt} \, 
\end{equation}
where the first
arrow refers to the incident photon helicity and the second to the
recoil proton helicity.
In the handbag mechanism, this is related
to the form factors by the expression \cite{Diehl} 
\begin{equation}
A_{LL}\frac{d\sigma}{d\sigma_{\rm KN}}\,=\,A_{LL}^{\rm KN} R_V(t) R_A(t) \, ,
\label{eq:all} 
\end{equation}
where $A_{LL}^{\rm KN}$ is the result for a structureless
Dirac particle.  This is plotted in Fig.~\ref{fig:pseudo}, where one sees
that the handbag calculation essentially follows the KN result, whereas
the pQCD prediction looks very different, thereby providing another stringent
test of the reaction mechanism in addition to a measurement of the axial
form factor $R_A$.  One can similarly define the transverse polarization
transver observable $A_{LT}$, which arises as an interference between proton
helicity flip and non-flip amplitudes.
In the strict pQCD limit, it must vanish since hadron
helicity is conserved.  Thus far this observable has not been
calculated with either the pQCD or handbag mechanisms.  However,
one can anticipate that in the handbag mechanism it will be
proportional to $R_T$, the RCS form factor that is closely related both
to the Pauli elastic form factor $F_2$ and to the quark total
angular momentum (see Table~1), both of which are topics of
high current interest.
The induced polarization $P_N$ is the component of recoil
polarization normal to the scattering plane and involves the
imaginary part of the interference between helicity flip and
nonflip amplitudes.  In the handbag mechanism, it is suppressed
since all amplitudes are strictly real in this model.\cite{Diehl}
In the strict pQCD limit, it vanishes due to hadron helicity
conservation.  No calculation has yet been done for this quantity.

\section{A New Experiment:  JLab E99-114} \label{sec:expt}
Experiment E97-108 (recently upgraded to E99-114) at JLab plans 
to measure differential cross sections for Compton scattering
from the proton at incident
photon energies between 3 and 6 GeV ($s$=6-12 GeV$^2$) and over a wide range of CM scattering
angles ($-t$=2-7 GeV$^2$).
The goal is to achieve a statistical precision of order 5\%,
with systematic errors less than 6\%.
In addition a measurement of the components of the proton recoil
polarization at $s=8$, $-t=4$~GeV$^2$ is planned, using
a polarized photon beam.  Both sets of
measurements utilize the technique shown schematically in
Fig.~\ref{fig:layout}.

A high duty factor electron beam with current $\ge$ 10 $\mu$A
is incident on a 6\% copper radiator
located just upstream of the scattering target.  The mixed beam of
electrons and bremsstrahlung photons is incident on a 15-cm LH$_2$ target.
For incident photons near the bremsstrahlung
endpoint, the recoil proton and scattered photon are detected with
high angular precision in a magnetic spectrometer and photon
spectrometer, respectively.
\piccaptioninside
\piccaption{\label{fig:layout}
Schematic of the planned JLab experiment.}
\parpic[r]{\fbox{\psfig{file=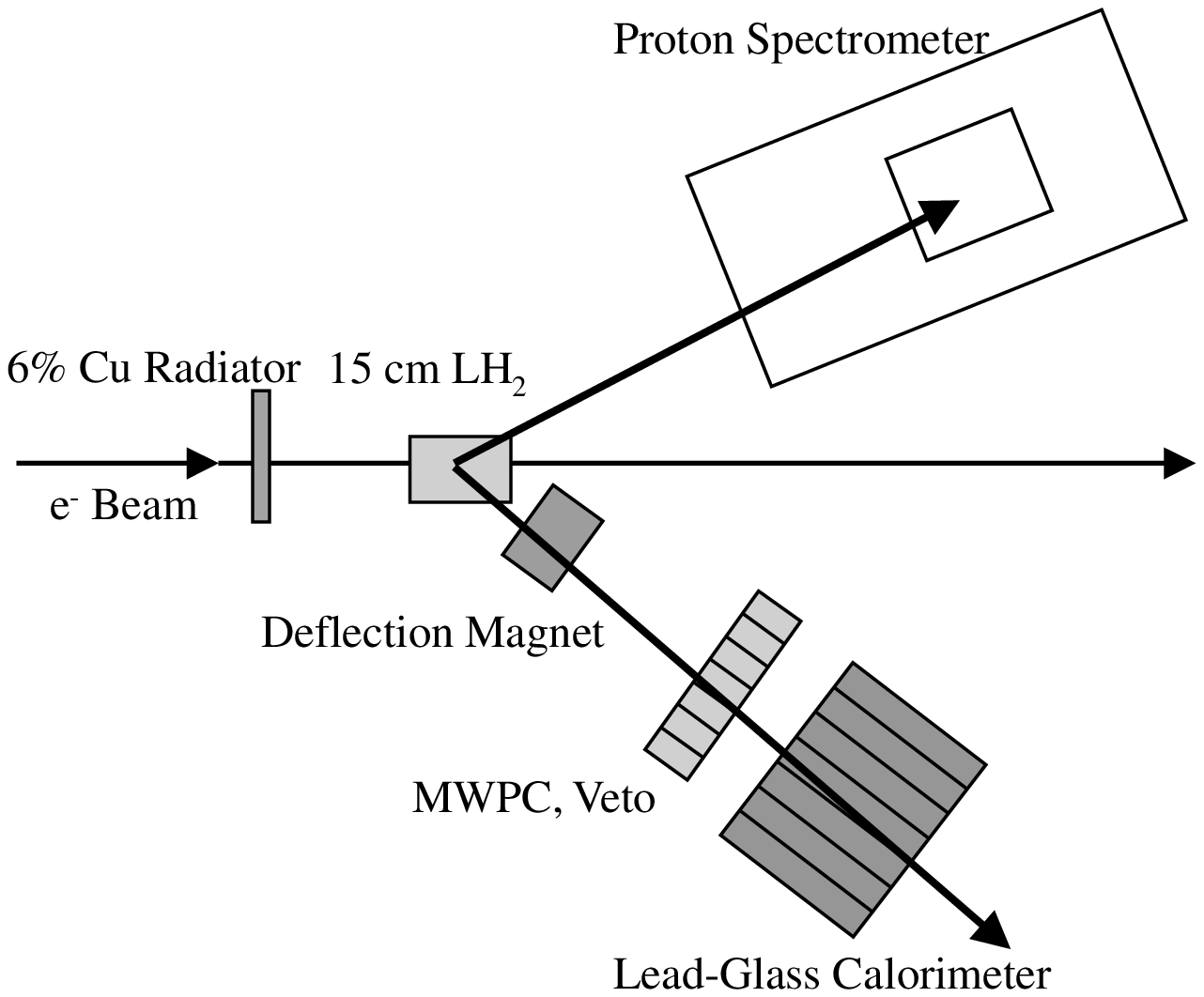,width=5cm}}}
\noindent
The magnetic spectrometer is one of the High Resolution
Spectrometers that are part of the standard Hall A equipment, along with
the cryogenic hydrogen target and bremsstrahlung radiator.  The
photon spectrometer is a new piece of equipment which is being constructed for
this experiment.  For the polarization measurements, a longitudinally
polarized electron beam is used and the polarization is nearly completely
transferred to the bremsstrahlung photon.  The components of the
polarization of the recoil proton are measured in the focal plane
polarimeter (FPP) which is also part of the standard Hall A equipment.
One essential feature of the experimental technique is the use of
the kinematic correlation between the scattered photon and recoil
proton in the RCS reaction to reduce the background of $\pi^0$
decay photons from the $p(\gamma,\pi^0 p)$ reaction, thereby
placing stringent demands on the combined angular resolution of
two-spectrometer system. A second essential feature is the mixed
electron-photon beam, which is required in order to achieve the
desired photon luminosity. On the one hand, this introduces the
necessity to identify and reject electrons from $ep$ elastic
scattering, while on the other hand it provides a convenient tool
for an {\it in situ} calibration of the photon spectrometer and
normalization of cross sections.

The principal new piece of equipment for this experiment is the
photon spectrometer, which is currently under construction.
The principal component
is a large-area segmented Pb-glass calorimeter with 
excellent position resolution and modest energy
resolution.  In order to reduce
the potential background of electrons from $ep$ elastic
scattering, which are kinematically indistinguishable from RCS
photons, several techniques will be used. First, the number of
$ep$ elastic electrons will be considerably reduced by avoiding
the region very close to the bremsstrahlung endpoint, where the
$e/\gamma$ ratio in the beam is very large.  Next, electrons will
be identified in a plexiglass {\v C}erenkov veto detector that is
segmented to allow for a veto that is spatially correlated with an
event in the calorimeter.  Finally, a magnet will be used to
deflect the $ep$ elastic electrons by a sufficient amount
on the front
face of the calorimeter to allow identification by the
altered kinematic correlation with the recoil proton relative
to undeflected RCS photons.  The mixed $\gamma-e$ beam is
advantageous in that the $ep$ elastic electrons can be used to calibrate
the photon spectrometer and normalize the RCS cross sections. 
For this purpose, several planes of MWPC
just in front of the calorimeter will be used in a separate {\it
in situ} $ep$ elastic scattering experiment to calibrate the
position of each element of the calorimeter and veto detector,
measure the position resolution, and measure the veto efficiency.
%The data acquisition electronics will utilize a variety of
%commercial NIM, CAMAC, and FASTBUS modules as well as some custom
%designed modules for the fast trigger. The entire spectrometer
%will be mounted on a mechanical assembly that allows changes in
%both scattering angle and radial distance, the latter needed to
%match the photon acceptance to that of the proton at different
%kinematic settings.

The technique outlined here is conceptually similar to that used
in the Cornell experiment.  However, the combined effects of a
high duty factor electron beam, a state-of-the-art
magnetic spectrometer, the ability to calibrate {\it in situ} with
$ep$ elastic scattering, and high segmentation in the photon
detector should allow significantly better measurements in the
range of $s$ and $t$ already covered by Cornell, as well as
significant extensions beyond that. The equipment would also be
suitable for measurements at higher energies, should those
energies become available at JLab in the future.

\section{Summary} \label{sec:summary}
In this contribution, I have summarized our present theoretical understanding
of the RCS process in the hard scattering regime.  In addition, I have presented
the conceptual design of a new experiment that should extend the RCS data
base in the very near future.

\section*{Acknowledgments}
It is a pleasure to acknowledge stimulating discussion with 
M. Diehl, P. Kroll, M. Vanderhaeghen, and especially
A. Radyushkin.  I also thank my E99-114
colleagues C. Hyde-Wright, B. Wojtsekhowski, and F. Sabatie.  
This work was supported
in part by the USNSF under Grant No. 94-20787.

\end{document}